\documentclass[preprint]{aastex}

\def\kms{\relax \ifmmode {\ \rm km s}^{-1}\else \ km\ s$^{-1}$\fi}
\def\degree{\mbox{$^{\circ}$}}

\def\cm3{${\rm cm}^{-3}~$}

\def\ha{H$\alpha~$}

\shorttitle{Kinematical analysis of BPNe}
\shortauthors{Dobrin\v{c}i\'{c}, et al.}

\usepackage{natbib}
\begin{document}

\title{Kinematical Analysis of a Sample of Bipolar Planetary Nebulae}

\author{Martina Dobrin\v{c}i\'{c}}
\affil{Instituto de Astrofis\'{i}ca de Canarias, 38200 La Laguna,
  Tenerife, Canary Islands, Spain; nina@iac.es}
\author{Eva Villaver\altaffilmark{1}}
\affil{Space Telescope Science Institute, 3700 San Martin Drive,
Baltimore, MD 21218, USA}
\author{Mart{\'{\i}}n A. Guerrero and}
\affil{Instituto de Astrof\'{\i}sica de Andaluc\'{\i}a,
Consejo Superior de Investigaciones Cient\'{i}ficas (CSIC), Apartado
Correos 3004, E-18080 Granada, Spain}
\author{Arturo Manchado\altaffilmark{2}}
\affil{Instituto de Astrofis\'{i}ca de Canarias, 38200 La Laguna,
  Tenerife, Canary Islands, Spain}

\altaffiltext{1}{Affiliated with the Hubble Space Telescope Space Department
  of ESA} 
\altaffiltext{2}{Affiliated Consejo Superior Investigaciones Cient\'{i}ficas,
  Spain}  
\begin{abstract}

We present the kinematics of a sample of bipolar planetary nebulae (PNe) 
which cover a wide range of observed morphologies and collimation degrees, 
from bipolar PNe with a marked equatorial ring and wide lobes to highly 
collimated objects.  
We use an empirical model in order to derive the expansion velocity, 
collimation degree, and inclination angle of the PN with respect to 
the plane of the sky. 
The equatorial expansion velocities measured in the 
objects in our sample are always in the low to medium range (3--16\kms),
while their polar expansion velocities range from low to very high 
(18--100\kms).  None of the objects in our sample, even those that show 
an extreme 
collimation degree, seem to be (kinematically) younger than $\simeq10^3$ yr.
We compare our results with the state-of-the-art theoretical models for the
formation of bipolar PNe. We find  good agreement between the
observed expansion velocities and numerical models that use magnetic 
fields with stellar rotation as collimation mechanism.  

\end{abstract}

\keywords{planetary nebulae: bipolar --- kinematic: planetary nebulae ---
  individual(Hen\,2-428, K\,3-58, M\,2-48, Hen\,2-437, K\,3-46, M\,3-55, 
  WeSb\,4, M\,1-75, and M\,4-14} 

\section{INTRODUCTION}

The interacting stellar winds model, initially proposed by 
\citet{1978ApJ...219L.125K}, is now the widely accepted 
scenario to explain planetary nebula (PN) formation.  
However, it has not yet been unequivocally determined what is the 
ultimate mechanism responsible for the collimation of the subclass 
of PNe that show bipolar or highly collimated morphologies 
(e.g. \citealt{2002ARA&A..40..439B}).  
In the context of the interacting stellar winds model, the formation 
of a bipolar PN requires a collimation mechanism. An isotropic fast wind
interacting with an aspherical mass loss structure formed during the top of
asymptotic giant branch (AGB) phase is the most commonly invoked scenario,
although
a collimated fast wind and a spherical AGB structure 
has also been considered \citep{2003ApJ...586..319L}. Despite the fact that
models are able to reproduce a wide variety of observed morphologies, it is
still not clear how the aspherical mass-loss during the AGB (or the
axisymmetric fast wind) is produced. 

From the numerical standpoint, several variations and additions to
the interacting stellar wind model have been developed extensively 
in the literature under the name of Generalized Interacting Stellar 
Winds (GISW) model
\citep{B87,1989AJ.....97..462I,1989ApJ...339..268S,Fm94,1995MNRAS.273..401M, 
1992A&A...253..224I,1994A&A...290..915M}.
In the GISW model, a fast, tenuous wind from the central star expands 
into a slow, dense wind whose geometry is assumed to be toroidal. 
Magnetic fields have also been considered  in 
numerical models; the magnetized wind blown bubble model (MWBB) produces 
an aspherical mass distribution by including toroidal magnetic fields that
constrain the outflow and produce jets in the polar direction
\citep{1999ApJ...517..767G}.  In addition, the precession of an episodic jet has been used
to reproduce point-symmetric morphologies   
\citep{1995ApJ...447L..49C,1998ApJ...508..696S,2001ApJ...557..256S,
2000ApJ...544..336G}. Variations to the magnetic
approach to generate the AGB aspherical density structure include using a
stellar companion rapidly rotating around the 
central star  \citep{1983RMxAA...5..319C,Gf04}.  
Although models accurately replicate the PNe shapes, the ultimate
question still remains: whether the magnetic 
field, and/or the stellar rotation, required to develop an aspherical
AGB mass-loss can be sustained by a single star or if they
require the presence of a binary companion \citep{N07}. 

In order to distinguish between the physical processes that may play a 
role in the process of shaping PNe,  it is important to have a 
detailed morphological classification scheme that includes the basic 
morphological features \citep{1992A&AS...96...23S,1997A&A...318..256G}, 
and also allows for more detailed subclasses (e.g. the presence of 
multiple shell PNe, multipolar axis) within each group 
\citep{2000ASPC..199...17M}. We should proceed further in the classification and explore the 
degree of collimation observed in bipolar objects, as this most likely 
reflects the type(s) and strength(s) of the physical process(es) 
involved. Finally, the kinematics of the nebula allows us to recover
its 3-D structure, as it gives access to the extra-dimension hidden
in direct imaging. 
 
In this paper, we present high resolution echelle long-slit spectroscopy of 
a sample of 9 PNe from the \citet{1996ApJ...466L..95M} catalog that show
highly axisymmetric morphologies with different degrees of 
collimation.   Seven of the PNe in our sample are classified as bipolar 
(K\,3-46, K\,3-58, Hen\,2-428, Hen\,2-437, M\,2-48, 
M\,3-55, and WeSb\,4), and the other two (M\,1-75 and 
M\,4-14) present quadrupolar morphologies that are 
characterized by two pairs of bipolar lobes symmetric with 
respect to two different axes.
Two objects in the sample, K\,3-58 and M\,4-14, show also point 
symmetric features.  
In \S2 we describe the observations and data reduction, 
in \S3 we outline the procedure used in the analysis of the data, and 
in \S4 we present the results of the kinematical fits performed to the objects in our sample. 
Finally, the results are discussed in \S5 and the conclusions summarized in \S6.

\section{OBSERVATIONS}
The images are from ``The IAC morphological Catalog of Northern Galactic
Planetary Nebulae''  \citep{Man96}. The long-slit echelle spectra were
obtained in July 1995  and August 1996  
at the 4.2m William Herschel Telescope (WHT) using the Utrecht Echelle 
Spectrograph (UES). 
A Tektronix CCD detector  1024$\times$1024 pixels, with a spatial scale 
of 0\farcs36~pixel$^{-1}$ was used. 
The 79 line mm$^{-1}$ echelle was centered on the \ha emission line, 
covering the spectral range from 6530 to 6600 {\AA} with a dispersion of 0.07
{\AA}~pixel$^{-1}$. 
The slit width was set to 1\farcs1, providing
a spectral resolution of 0.14 {\AA}  which corresponds to 6.5 \kms. 
The unvignetted field of view is up to 160\arcsec~long. 

The spectra were reduced using 
standard IRAF tasks for two-dimensional spectra. 
The images were corrected for differential illumination of the slit using 
sky flat fields obtained with the same configuration.  
The wavelength scale and geometrical distortion were set by a 
two-dimensional fit to a Th-Ar calibration lamp. 
The wavelength calibration has an accuracy
better than 0.004 {\AA} (0.2 \kms). 
The sky emission, which includes the geocoronal \ha line, 
 was removed by fitting and subtracting the background using a 
low order polynomial function.

The observing log is summarized in Table~\ref{observations} were the common 
PN name is given in column (1), the PN~G number in column (2), the date of 
the observations in column (3), the slit position angle (P.A.) in column 
(4), and the exposure time in column (5). 
The slit positions 
go through the center of the nebulae, and their P.A.\ have been chosen 
to cut across their main symmetry axis.
In most cases, spectra were obtained at two slit positions along 
perpendicular directions.

\section{FITTING PROCEDURE}

Bipolar PNe have typically hourglass or butterfly morphology, 
meaning that they show two bipolar lobes connected by a 
narrow waist.  
The exact geometry can be very different from one bipolar PN to another.  
These differences can basically be described by the aspect ratio of the 
bipolar lobes and by the relative width of the waist with respect to 
these bipolar lobes.  
In addition, the inclination with respect to the plane of the sky affects 
the apparent morphology, while the nebular kinematics 
determines the observed expansion velocity.  
In order to determine the 3D geometry and kinematics  
of the bipolar PNe in our sample, we have used the empirical model 
of \citet{1985A&A...148..274S} which was originally developed to 
interpret the velocity field of the bipolar nebula around the 
symbiotic star R~Aquarii. 
In this model, hereafter referred to as the Solf \& Ulrich's model, the 
velocity distribution along a plane going through the main nebular symmetry 
axis is described by 
\begin{equation} \label{solf}
v(\alpha) = v_{\rm e} + (v_{\rm p} - v_{\rm e}) \sin^{\gamma}(\arrowvert \alpha
\arrowvert)
\end{equation}
\noindent
where $\alpha$ is the latitudinal angle from the symmetry axis 
of the nebula, $v_{e}$ is the equatorial velocity, 
the minimum velocity along the nebular waist,  $v_{p}$ is the polar velocity, 
the maximum velocity of the farthest point from the center of the nebula,
and $\gamma$ is a geometrical factor that defines the hourglass geometry, 
with $\gamma\approx1$ for round, bubble-like bipolar lobes, and 
$\gamma>>1$ for very elongated bipolar lobes.

One advantage of the Solf \& Ulrich's model 
is the possibility to fit simultaneously 
images and spectra of a given nebula in order to recover the 
3D geometry lost by projection effects.  
As an illustration, we show in Fig.~\ref{fig4inc.ps} simulated 
images and position-velocity (PV) diagrams of a nebula 
obtained using Eq.~\ref{solf} for three different inclination 
angles with respect to the plane of the sky.   
From Fig.~\ref{fig4inc.ps}, it is clear 
that the tilt of the simulated line strongly depends on the inclination angle 
of the object. 
Therefore, the shape of the PV diagram can be used to derive 
the inclination angle which may only be assessed from the 
aspect ratio of the central ring under the strong assumption that it 
is circular.
Similarly, Fig.~\ref{fig3gamma.ps} shows simulated images and PV 
diagrams for nebulae with different $\gamma$ factors. 
In this case, we see that as $\gamma$ increases, the shape of the PV 
diagram becomes narrower and the line tilt more apparent. 
Fig.~\ref{fig3gamma.ps} also shows that it is not possible to determine 
the inclination angle of the nebula from the projected shape of its 
cylindrical narrow waist for $\gamma\ge9$.  
However, it is readily seen that we can determine the inclination from 
the line tilt in the PV diagram. 

The fitting procedure is an iterative process that is started by 
modeling simulated images and PV diagrams with parameters that 
are first order estimates. 
The set of free parameters includes the polar and equatorial velocities, 
$v_{\rm p}$ and $v_{\rm e}$, the 1 kpc kinematical age \footnote{
We use the term 1 kpc kinematical age throughout the text to refer 
to the age obtained from the kinematical fitting assuming that the nebula is
located at a distance of 1 kpc. The 1 kpc kinematical age that is determined in the model
explicitly assumes that the lobes and the waist started to grow
simultaneously. The {\it distance-corrected kinematical age} is the 1 kpc age 
corrected for the distance to the object.},  
the inclination angle of the PN with respect to the plane of sky,
the $\gamma$-factor describing the shape of the nebula, and the radial 
velocity.  

For the objects with a clearly visible central ring structure, we can 
estimate both the equatorial expansion velocity and the radial velocity 
just by measuring the velocity of the two brightness maxima at the 
opposite sides of the ring; 
the average of the two velocities gives us the radial velocity, while the 
half difference of the velocities give us a lower limit of the equatorial 
expansion velocity. 
In this case, we can even assess the inclination of the nebula 
as the axis ratio of the equatorial ring is a direct measurement of 
the inclination angle if we assume that the ring is circular. 
Then, the synthetic nebular image and PV diagram are plotted over the
H$\alpha$ or H$\alpha$+[N~{\sc ii}]  
image and [N~{\sc ii}] spectra\footnote{
For the spectral fits, we use the [N~{\sc ii}] emission line 
because of its relative brightness and small thermal broadening.}, 
respectively.  
The initial estimates of the parameters are changed until we reach 
an optimum fit when the synthetic PV diagram passes through the 
intensity maxima in the spectral data and the simulated image traces tightly 
the nebular geometry.
 
We note that the geometry of the nebula can be reproduced with different 
sets of values for the free parameters.
However, the degeneracy is broken by fitting 
the geometry of the object together with the spectral line shape.  The
accuracy of this iterative process is hard to quantify and it has to be
determined for each object individually. 

The parameters of the best fits are listed in Table~2 and 
Figs.~\ref{double1} to \ref{double6} show the images and PV diagrams 
overplotted with our best fits to the data. The velocities in the horizontal
axis of Figs.~\ref{double1} to  
\ref{double6} have been converted to the LSR system. 
The images are shown at the same scale as the spectra and have been rotated 
to allow a direct comparison of the kinematical features along the spatial 
direction.

\section{RESULTS:  THE FITS TO THE IMAGES AND SPECTRA}

In the following we provide a detailed description of both the data and the 
best fit parameters obtained for each object.

\noindent
{\bf Hen\,2-428}\\
Hen~2-428 (Fig. \ref{double1}) is a bipolar PN with a noticeable 
equatorial ring and two open hourglass bipolar lobes of which the Northern 
one is brighter and more extended.
The central star, unusually bright for bipolar PNe, has a known binary 
companion \citep{2001A&A...377.1042R}. 
For this nebula, we acquired 
long-slit echelle spectra along 
PA's, 77\degree\ and 157\degree\ (see Fig. \ref{double1}). 
We note that the spectrum taken along P.A.~77\degree\ is off-center and 
passes through the bright knots in the equatorial ring.
Therefore, the basic spectral fitting relies on the spectrum at 
P.A.=157\degree\ which has adequate signal-to-noise ratio in the central region and the 
brightest Northern lobe to constrain the fit.  
From our best fit, we get an equatorial expansion velocity, $v_{\rm e}$, of 
16\,\kms, and a polar velocity, $v_{\rm p}$, of 80\,\kms.  
A value of $v_{\rm e}$=15\,\kms\ was obtained for this object by 
\citet{2001A&A...377.1042R}.

\noindent
{\bf Hen\,2-437}\\
Hen\,2-437 (see Fig.\ref{double2} [Top]) is a very elongated PN 
that does not show a central ring structure.  
These morphological features make
very difficult to constrain its equatorial velocity and to 
reproduce the spectral shape with the Solf \& Ulrich's model.
Since there is almost no tilt in the central maxima of the [N~{\sc ii}] echellogram 
along PA~77\degree, we assume that the nebula has a small 
inclination angle, lying almost in the plane of the sky.

The geometry of the nebula is thus constrained by its morphology, and 
the nebular contours can be reproduced relatively well by using an extremely 
high $\gamma$ factor.
Given the difficulties in reproducing the spectral shape of Hen\,2-437, 
it was not possible to derive firm results.  
We give a set of the most likely values: the inclination angle is 
small, within 2\degree, the polar velocity can be constrained to 
be between 50 and 100 \kms, and the equatorial velocity
has to be small, of the order or lower than 10\kms.

\noindent
{\bf K\,3-46}\\
K~3-46  (Fig.~\ref{double2} [Bottom]) has a well defined hourglass geometry 
with a prominent equatorial ring and a marked waist.  
For this nebula, we obtained very low expansion velocities 
from the fit to the PV diagram, with $v_{\rm e}<3$\,\kms and 
$v_{\rm p}\sim18$\,\kms. We note that it was not possible to find an optimal
fit, and 
Fig.~\ref{double2} [Bottom] indeed suggests that our simulated fit 
would require a larger central ring.  
We also note that the kinematical age of K\,3-46 is of 24,000 yr 
which is, as we will discuss later, the highest value obtained 
in our sample. 

\noindent
{\bf K\,3-58} \\
At a first glance, K\,3-58 shows a classical bipolar morphology
(Fig.\ref{double3}), but its conspicuous ring is rather irregular 
and the bipolar lobes show point-symmetry. 
All these components are clearly detected in the [N~{\sc ii}] 
echellogram obtained along P.A.~90\degree, with the brightest emission peaks 
located on the equatorial ring.
Inside the main lobes, a secondary structure reaches approximately up 
to one half of the length of the main lobes.
This secondary structure, that is clearly present in the spectral data 
as bright emission peaks, may have originated from a later ejection to 
the one that shaped the main lobes. 
From the fit we obtain $v_{\rm p}$=38\kms\ and $v_{\rm e}$=12\kms.  
Interestingly, the ratio between the equatorial and the polar velocities 
is relatively small for this object.

\noindent 
{\bf M\,1-75}\\
M\,1-75 (Fig.~\ref{double4})
is the first object classified as
quadrupolar \citep{1996ApJ...466L..95M}. 
M\,1-75 with its complex morphology and extremely high He abundances 
and N/O ratio is a candidate for a peculiar stellar evolutionary path 
(Guerrero et al.\ 1996). 
The nebula does not show a circular equatorial ring, but an 
irregular annular structure at the central regions.  
Since the equatorial ring is clearly not circular, it cannot be used 
to assess the inclination angle of the nebula with respect to the plane of 
the sky.

The spectra for M\,1-75 was obtained along the main symmetry axis of the two 
pairs of bipolar lobes, namely along P.A. 150\degree\ and 176\degree.  
The former spectrum (Fig.~\ref{double4} [Top]) proved more useful 
to guide our spectral fit, as it provides information for both pairs of 
bipolar lobes. The data taken at the second position angle has been 
used for control purposes only and is shown at the bottom of Fig.~\ref{double4}.

The large lobes, along P.A. 150\degree, can be well fitted by 
using $\gamma$=6 and an inclination angle of $87\degree$ as determined 
from the spectra.  Since these lobes seem to lie close to the plane of the
sky, a change in just 2\degree\ is very noticeable for such an elongated
shape, and the inclination angle can be constrained well. The expansion
velocities obtained for the largest pair of lobes are relatively low, with
$v_{\rm e}$=8\,\kms\ and $v_{\rm p}$=55\,\kms.   

The kinematics of the small lobes cannot be constrained as 
accurately because their geometry in the image is not well defined.
We have assumed, for simplicity, that the small and large pairs of lobes
have  
identical morphologies, and therefore we use the 
$\gamma$ factor determined for the large lobes, $\gamma$=6. 
The range of inclination angles we then obtain is  between 60\degree\ and  
70\degree.  Using an inclination angle of 65\degree, we obtain 
expansion velocities of $v_{\rm p}$=45\,\kms\ and $v_{\rm e}$=12\,\kms.

\noindent
{\bf M\,2-48}\\
M\,2-48 (see Fig. \ref{double5} [Top]) does not show an equatorial ring, 
but a pinched waist.  
The pair of collimated lobes ends in a pair of bow-shocks that
were analyzed in detail by \cite{2002A&A...388..652L}. 
The collimated morphology of the bipolar lobes of M\,2-48 requires 
to use a high $\gamma$ factor ($\sim 8$). The narrow central waist of M\,2-48
does not allow to determine the  
inclination angle from the geometry of the ring and, therefore, as 
mentioned in \S3, we have to rely on the line tilt in the observed 
PV diagram.
We reproduce well the geometry of both image and spectra by using an
inclination angle of $\sim80\degree$, with the North-Eastern lobe 
receding from us.  
We derive a radial velocity of 16\,\kms, a $v_{\rm e}$=10 \kms, and a 
$v_{\rm p}$ of 100\,\kms\ which is the largest value in our sample. 
The inclination angle and expansion velocities obtained are consistent 
with the determination by \citet{2002A&A...388..652L}.  

\noindent
{\bf M\,3-55} \\
Despite being very faint,  M\,3-55 displays a clear symmetric shape 
(Fig.~\ref{double5} [Bottom]). 
We obtain for M\,3-55 a low expansion velocity, 
lower than 6\,\kms\ in the equator and about 19.5\,\kms\ 
along the polar directions.  
M\,3-55 also has the lowest $\gamma$-factor of all the objects within 
our sample which is not surprising given that it has a relatively
wide ring and round lobes.  

\noindent
{\bf M\,4-14} \\
\cite{1996ApJ...466L..95M} classified M\,4-14 (Fig.~\ref{double6} 
[Top]) as a PN  with quadrupolar morphology (based on imaging). 
Here, we also remark the noticeable point-symmetry of its 
bipolar lobes. To fit the cylindrical waist of M\,4-14, a high $\gamma$ factor 
($\gamma$=5) is needed. From this fit we obtain $v_{\rm e}=11$\,\kms\ and
$v_{\rm p}=$65\,\kms. 

M\,4-14 has a [N~{\sc ii}]/\ha ratio among the highest found in PNe, 
indicative of a high nitrogen enhancement and high N/O ratio, that are 
known to be correlated with bipolarity \citep{1983IAUS..103..233P}. 
The chemical abundances, expansion velocity, and geometry factor 
$\gamma$ make M\,1-14 and M\,1-75 very similar. 

\noindent
{\bf WeSb\,4}\\
WeSb\,4 is a large, somehow diluted object, that shows one of the 
most irregular morphologies among the objects in the sample 
(Fig.\,\ref{double6} [Bottom]). 
The narrow-band image does not display a clear hourglass shape that, 
on the other hand, is evident in the much deeper spectrum. In the spectrum 
we detected weak, and more extended emission  than in the 
optical images.  
From the image and [N~{\sc ii}] echellograms 
along P. A. 69\arcdeg\ and {159\arcdeg}, we  derive an inclination 
angle of 50\degree, and a $v_{\rm e}$ of 14\,\kms, and $v_{\rm p}$ of 
95\,\kms\. WeSb\,4 has one the highest $v_{\rm p}$ in our sample.  

\section{DISCUSSION}

We summarize in Table~\ref{results} the parameters obtained from the
best fits to the data.  Column (2) gives
the size of the lobes and waist as measured from the maximum extension of
the 3$\sigma$ contour levels extracted from the images obtained by Manchado
et al. (1996). The polar and equatorial expansion velocities are given in column (3) and 
(4), respectively, the 1~kpc kinematical age is given in column (5), 
the inclination angle in column (6), and the $\gamma$ factor in column 
(7).  Below the values obtained from our best fit (when the synthetic PV
diagram passes through the intensity maxima in the spectral data and the
simulated image traces tightly the nebular geometry), we have listed the range 
of values that still provide reasonable matches (when we can still provide a fit
that passes through the main features) to the images and 
spectra in order to provide an estimate of the uncertainties in our 
fits. 

The equatorial velocities 
range from very low values (3\,\kms~for K\,3-46) to typical 
expansion velocities (16\,\kms~for Hen\,2-428).  
None of the objects in our sample has high, $\sim$40\kms, equatorial
expansion velocities such as the  
ones found by \citet{1993A&A...278..247C} in the bipolar PNe CTS\,1 
and Hen\,2-84.   
The polar velocities cover the whole range, from very low (18\kms\ in K\,3-46) to medium/high 
(100\kms\ in M\,2-48). 

The PNe analyzed in this paper can be classified into three different 
groups, according to their morphology and kinematics.  
The first group, formed by Hen\,2-428, K\,3-46, K\,3-58, M\,3-55, and 
M\,4-14 consists of bipolar PNe with notable central rings.  
For these objects, we have a direct estimate of their equatorial 
expansion velocities measured from the spectral emission of the 
ring (as it was described in \S3). 
A second group is formed by the highly collimated objects 
Hen\,2-437 and M\,2-48 having the highest polar expansions and elongated
bipolar morphology typical mostly for younger PNe. 
Finally, the third group is formed by the somewhat deteriorated
objects with  
poorly defined morphologies; M\,1-75 and WeSb\,4. 
The PNe belonging to this third group are expected to be more evolved 
since they have the highest kinematical ages.  
Although K\,3-46 also has a large kinematical age, we did not include 
it in this group of evolved objects.  
We suspect that its 
large kinematical age is probably a consequence of deceleration in the course of its 
evolution.

\subsection{PN Ages}

Our spatio-kinematical study provides a direct estimate of the 
kinematical age that can be used to assess the nebular age.  
It is important to keep in mind in this comparison the very likely 
possibility of acceleration or deceleration of the nebular material 
due to the complex interaction between the ionization and dynamics 
of the shell driven by the hot bubble.  
As a result, kinematical ages often do not match the age of the central 
stars, as shown theoretically by \cite{2002ApJ...581.1204V}, who proved 
that kinematical ages overestimate the age in young nebulae and 
underestimate it for evolved ones.  Moreover, the definition of PN age is a rather
tricky concept. It can be considered that a PN is born when the central star
supplies enough photons capable of ionizing the nebula. However, the kinematical
age is a dynamical concept that tell us when the gas started moving. At the
time of nebular ionization the gas is already moving  
and the fotoionization itself is expected to change the gas dynamics. In
addition, the gas velocity as
inferred from models \cite{2002ApJ...581.1204V} is not constant. If the gas
has been accelerated the kinematical ages are underestimated and the opposite
is true if the gas has suffered deceleration.
The age obtained in Table~2 is one of the
parameters of the fitting and it is obtained under very simplistic
assumptions, namely that the lobes and the waist were formed at the same 
time and they have been moving at a constant velocity since then. Therefore, the ages
determined this way are just an order of magnitude approximation to the time
since the formation of the nebula (both dynamically and from
ionization).  

The kinematical model assumes the same age for the
lobes and the waist of a nebula. As a consequence the larger lobes always
have larger expansion velocities. Note that  under this simplistic assumption we are
excluding the possibility that the lobes were formed before the waist.

We provide in Table~\ref{results} 
the 1~kpc kinematical ages.
Here, we have estimated the ``real'' ages by multiplying the 1~kpc 
kinematical ages derived from our fits by the individual distances.  
It is well known that PNe distances
are poorly determined, however they are a necessary parameter 
for the age estimation. It is important to note that the ages determined in
Table~3 are the ones given in Table~2 but scaled to the distance to the nebula.
We have used the distances obtained 
from the Galactic rotation curves of Burton (1974) when possible, 
as these distances are considered to be the most reliable.  
In some cases, our data was out of the range covered by the rotational 
curves, and therefore the distance estimate was not viable through this 
method.  
This also applies for M\,2-48, for which this method gives ambiguous results, 
and for M\,1-75, for which the distance derived from this method (5.4 Kpc) is
unreasonable and gives a too long kinematic age for a PN.
When it was not possible to estimate the distance from the rotational curves, we have used the 
statistical distances from Cahn et al (1991) when available.  
Otherwise we have used the distance from \citet{1971ApJS...22..319C}
or, as a last resort, an average of available distances in the literature.
We note that the values given by \citet{1984A&AS...55..253M} are 
systematically lower than the values given by others.  
Therefore, Maciel's distances have not been used for 
the averages.

In Table \ref{ages}, 
column (1) gives the PN name, 
column (2) the radial velocity, 
column (3) the kinematic age at 1~kpc obtained from the model fitting, 
column (4) 
the statistical distances taken from \citet{1992secg.book.....A}, 
column (5) the distances estimated from the galactic rotation curve 
(Burton 1974), and column (6) the distance used to estimate the kinematical age which is given in 
column (7).

The distance-corrected kinematical ages range between $\sim$5,000 and 
$\sim$20,000\,yr. 
Intuitively, it is expected that younger PNe would have better defined
morphologies than the older ones.  
 This appears to be the case for almost all the objects analyzed. 
Hen\,2-428, M\,2-48, M\,3-55, and M\,4-14 are relatively young, 
with kinematical ages $\sim$5,000 \,yr, and all have 
well defined morphologies. 
M\,1-75 and especially WeSb\,4 are older, with kinematical ages 
$\sim$10,000 \,yr, and their morphologies are not sharp.

In old objects such as WeSb \,4, ionization-driven instabilities 
which act on a timescale comparable to the kinematical age of the 
nebula might be responsible for the development of irregular shapes 
\citep{1995ApJ...447L..49C}.

There are two objects in our sample for which a relation between their age
and their morphology is not 
straightforward. 
K\,3-46, the oldest PN in our sample ($\sim$20,000 yr), and K\,3-58, a 
relatively old PN, have sharp morphologies. 
One possibility is that the distances used are wrong, which would not 
be a surprise given the high uncertainties involved in PN distance 
determinations.  
Another possibility for K\,3-46 is that it has experienced deceleration, 
in which case the age given in the Table~3 would be an underestimation 
of the real age of the object.  

Adopting 
a distance of 1\,kpc to Hen 2-437, this is the youngest 
PN in our sample, with a kinematical age between 750 and 2000\,yr.
This age is consistent with those of high collimated PNe 
\citep{2003ApJ...586..319L}.  

\subsection{Comparison with Numerical Models}

Quite often the comparison between theoretical models and 
observed objects is based only on a morphological match since this 
the only available information in most cases.
From our experience in matching published models to data, we have
found that it is possible to find models that can simulate the
morphology of a PN but with velocities  different to those 
measured; in this case we need to know whether the model can 
simultaneously reproduce the morphology and kinematical 
properties that, in the case of bipolar PNe, are parameterized by the 
expansion velocity, the physical nebular size, and geometry (the 
$\gamma$ factor). Note that, in the model of Solf \& Ulrich, the expansion of a 
bipolar PN is homologous, i.e., in its expansion, the nebula keeps 
its proportions, as well as the ratio between the polar and equatorial 
velocities.

Only the models by \citet{1999ApJ...517..767G} and \citet{2000ApJ...544..336G}
provide enough information on the evolution of the morphology 
for different collimation parameters for us to compare to our data.  
In the \citet{1999ApJ...517..767G} models (hereafter, GS99), the 
velocity ratio between the equator and the poles depends on the shaping
mechanism, while the absolute values of the velocities  
depend on the initial velocity of the slow wind.
Therefore, when evaluating the ability of the theoretical models to match our 
observations, we are more interested in comparing the nebular shapes 
and velocity ratios rather than the absolute velocity values.

To compare the models with the observations, we need a parameter that can 
quantify the degree of collimation.  The $\gamma$ factor in the Solf \&
Ulrich geometrical model is an appropriate parameter to account for the degree 
of collimation observed. The shape in the GS99 models 
is controled using different values of $\sigma$ (the ratio of the magnetic 
to the kinetic energy density in the fast wind) 
and $\Omega$ (ratio of the stellar rotational velocity to the 
critical breakup velocity). 

In order to allow a better comparison between
the models and the parameters derived from the observations we have determined
the Solf \& Ulrich's parameters ($\gamma$ factor) that
correspong to the GS99 simulated cases. We find that low $\sigma$ and $\Omega$ values 
produce shapes that are similar to those generated by 
low values of $\gamma$ ($\gamma<1$) (see  Fig.\ 5 in GS99).  
The GS99 cases with low $\sigma$ and high $\Omega$ values are equivalent 
to medium $\gamma$ values,  
while high $\sigma$ and high $\Omega$ values result in very collimated 
objects, with high $\gamma$ values. 

For categories J--K in GS99, the morphology develops using only rapid star 
rotation rates, while for categories Q--V the additional help of magnetic fields
is required. Hen\,2-428, K\,3-46, and M\,3-55, which have the lowest $\gamma$ 
(1, 0.9, 0.6, respectively) factors in our group, compare 
well to the categories J--K of GS99.  
Both the observed objects and the corresponding theoretical simulations 
are reproduced with $\gamma$ factors around 1 and, although our PNe do 
not have the same absolute expansion velocities as the GS99 
cases, this is not a critical issue 
as lower expansion velocities in GS99 magneto-hydrodynamic 
(MHD) simulations could be obtained by choosing lower initial velocities for the slow 
wind.  In fact, the ratio of expansion velocities for K\,3-46 and Hen\,2-428 
situates them in categories Q--R, but their $\gamma$ factor and 
morphology place them into the J--K group.  
It must be noticed that the central star of Hen\,2-428 is in a binary 
system (Rodr\'\i guez et al.\ 2001).  
It is tantalizing to consider that the presence of the binary 
companion may have influenced the nebular shaping, resulting 
in a bipolar PN with enhanced polar velocities with respect to 
the bipolar PN that would have been collimated by a single, 
rotating star.

In general, the expansion velocities in GS99 are rather high
compared to those observed in our sample. However, M\,1-75, M\,4-14,
 and WeSb\,4 have similar expansion velocity ratios (i.e. equatorial to polar 
expansion velocity) and morphologies as the PNe in categories Q--R in
GS99. These 3 PNe have $\gamma$ factors between 5 and 6. In fact, 
M\,4-14 even coincides in its kinematical age and 
value of its expansion velocities with the example given by GS99.

M\,2-48 is more collimated than the previous objects and fits categories 
S--T, while Hen\,2-437 would go even higher, to the class U in GS99. 
Such highly collimated objects, with lobes of wedge-shaped polar 
regions, are not well reproduced by the Solf \& Ulrich's model and therefore
our derived velocities carry a large uncertainty. 

Careful inspection of the images of M 2-48 and Hen 2-437, which have 
the highest collimation factors in our sample,  reveals traces of 
a dusty disk in the equatorial region, a feature not visible in the 
low-$\gamma$ objects.  
Similar features were already detected in various young, highly collimated 
objects (e.g. \citealt{2006ApJ...653.1241S}) and suggest the existence 
of thick equatorial disks.
 
Finally, we are left with K\,3-58 a rather peculiar object since it has 
an expansion ratio and $\gamma$ factor that lie outside of the cases 
modeled by GS99.  
K\,3-58 presents a cylindrical equatorial ring which is wider than the 
one obtained for the J--L categories of GS99, and it shows lobes 
that are rounder than those in the R--V classes. The
best fit to the morphology and kinematics of this object can be found in  
Fig.\,5 of \citet{2000ApJ...544..336G}. 
The nebula simulated there has almost identical expansion velocities 
to those we measure. In this model, the point-symmetric shape is accomplished
with a tilt between the  
magnetic collimation axis and the bipolar wind outflow.  

To summarize, we find that objects with low $\gamma$ values 
generally compare well to the GS99 models which are mainly
shaped by stellar rotation. 
The possible exceptions are PNe formed in binary systems that will present 
low $\gamma$ but somewhat higher expansion velocities (Hen\,2-428), 
while high $\gamma$ objects show high polar expansions and agree well with
the simulations obtained using strong magnetic fields and high rotational velocities.  

\section{CONCLUSIONS}

Although all of our objects are bipolar/quadrupolar, they differ
significantly in their morphology and kinematics. 
Morphologies range from 
bubble-lobed (Hen\,2-428) to highly collimated (Hen\,2-437) PNe.
The sharpness of the nebular shapes also varies from well defined 
objects to these with somewhat deteriorated shapes (WeSb\,4). 
This variety is reflected in the range of geometrical $\gamma$ factors 
(from 0.6 to 20) which might differentiate between the physical processes that
originate them. 

We find that  the objects from our sample present as well a variety of 
expansion velocities, from low (3\kms) to medium (16\kms) equatorial 
expansions, and from low (18\kms) to medium/high (100\kms) polar 
expansion velocities. The disagreement between the spectral data of K\,3-46, 
hardly revealing signs of equatorial expansion, and its image, showing a wide 
central ring indicate the possibility of deceleration in this nebula.
This deceleration may reveal the interaction of the nebular material 
with a dense equatorial disk.

The estimates of kinematical ages, derived using distances inferred from the
Galactic rotation curve or otherwise statistical distances, vary from middle age to old with possible 
significant errors originated from distance errors and non-uniform 
expansions.  
The data agree rather well with the state-of-the-art theory for PN
collimation, however, we cannot exclude the origin of the possible shaping mechanisms i.e. 
whether rapid star rotation, and/or magnetic fields are originated by single or binary systems. 
We suggest that the $\gamma$ factor used to fit the Solf \& Ulrich's model 
could roughly indicate which shaping process is actually 
at work.

\textbf{Acknowledgments:} 
M.D. thanks to Katrina Exter for her corrections and useful discussions.
The 4.2 m William Herschel Telescope and the 2.5 m Nordic Optical Telescope
are operated on the island of La Palma by the Royal Greenwich Observatory and
the Lund Observatory, respectively, at the Spanish Observatorio del Roque de
los Muchachos of the Instituto de Astrof\'{\i}sica de Canarias.  M. D. and
A. M. acknowledge support from grant \emph{AYA2004-3136} and M.A.G. from 
grant \emph{AYA2005-01495}  from the Spanish Ministerio de Educaci\'on y Ciencia.

\clearpage

\begin{figure}
\caption[1]{
 Synthetic images and 
position-velocity (PV) diagrams for different inclination 
 angles for a bipolar PN with $\gamma$=1.}
\label{fig4inc.ps}
\end{figure}


\begin{figure}
\caption[2]{ Synthetic images and position-velocity (PV) diagrams for a bipolar PN at an 
inclination angle of 30\degree\ with different values of $\gamma$.}  
\label{fig3gamma.ps}
\end{figure}


\begin{figure}
\centering
\protect
\caption[ ]{Kinematical fitting of the Hen\,2-428 for a slit position angle 
of 167\degree\ (\emph{top}) and for a slit position angle of 77\degree\ 
(\emph{bottom}). 
The narrow-band image, extracted from the \citet{1996ApJ...466L..95M} 
catalogue, has been rotated to the position angle of the slits along which 
the spectra were obtained.  
The angular scale of images and spectra have been matched to allow an easy 
comparison of images and spectra.  
Also note that the PV diagrams have been corrected for the local standard 
of rest (LSR) velocity. 
In order to display the complete structure of the spectral line, the 
contrast of the spectra is adjusted, so the faintest parts can be seen. 
}
\label{double1}
\end{figure}


\begin{figure}
\centering
\protect
\caption[ ]{
Same as Fig.~3 for Hen\,2-437 for a slit position angle of 77\degree\ 
(\emph{top}) and K\,3-46 for a slit position angle of 106\degree\ 
(\emph{bottom}).  
In the later, the central part of the spectral line is shown 
saturated, and white contours have been used to show levels of the 
same intensity.}
\label{double2}
\end{figure}


\begin{figure}
\centering
\protect
\caption[ ]{
Same as Fig.~3 for K\,3-58 for a slit position angle of 90\arcdeg\ 
(\emph{top}) and for a slit position angle of 14\arcdeg\ (\emph{bottom}). 
As for K\,3-46, white contours are used for the brightest regions 
of the spectral line. 
 }
\label{double3}
\end{figure}


\begin{figure}
\centering
\protect
\caption[ ]{
Same as Fig.~3 for M\,1-75 for a slit position angle of 150\arcdeg\ 
(\emph{top}) and for a slit position angle of 176\arcdeg\ (\emph{bottom}). 
 As for K\,3-46, white contours are used for the brightest regions 
of the spectral line. }
\label{double4}
\end{figure}


\begin{figure}
\centering
\protect
\caption[ ]{
Same as Fig.~3 for M\,2-48 for a slit position angle of 66\arcdeg\ 
(\emph{top}) and M\,3-55 for a slit position angle of 57\arcdeg\ 
(\emph{bottom}). 
 As for K\,3-46, white contours are used for the brightest regions 
of the spectral line.  }
\label{double5}
\end{figure}


\begin{figure}
\centering
\protect
\caption[ ]{
Same as Fig.~3 for M\,4-14 for a slit position angle of 40\arcdeg\ 
(\emph{top}) and WeSb\,4 for a slit position angle of 159\arcdeg\ 
(\emph{bottom}). 
 As for K\,3-46, white contours are used for the brightest regions 
of the spectral line of M\,4-14. } 
\label{double6}
\end{figure}

\clearpage

\begin{deluxetable}
{lcccc}
\tabletypesize{\scriptsize}
\tablenum{1}
\tablewidth{0pt}
\tablecaption{Spectroscopic Observations\label{observations}}
\tablehead{
\multicolumn{1}{c}{PN NAME} &
\multicolumn{1}{c}{PN G Number} & 
\multicolumn{1}{c}{Date} & 
\multicolumn{1}{c}{
$\begin{array}{c}
{\rm Slit P.A.}\\
(\degree)\\
\end{array}$} &
\multicolumn{1}{c}{$\begin{array}{c}
{\rm Exposure Time}\\
{\rm (s)}\\
\end{array} $ }}
\startdata
Hen~2-428 & 49.41$+$2.48 & 1995 Jul 15 & 77        & 1800 \\
          &              &             & 167       & 1800 \\
Hen~2-437 & 61.40$+$3.62 & 1995 Jul 15 & 77        & 1800 \\
          &              &             & 167       & 1800 \\
K~3-46    & 69.21$+$3.81 & 1995 Jul 16 & 16        & 1800 \\
          &              &             & 106       & 1800 \\
K~3-58    & 69.68$-$3.94 & 1996 Aug 07 & 14        & 900  \\
          &              &             & 90        & 1800 \\
M~1-75    & 68.86$-$0.04 & 1996 Aug 07 & 150       & 1800 \\
          &              &             & 176       & 1800 \\
M~2-48    & 62.49$-$0.27 & 1996 Aug 07 & 66        & 1800 \\
M~3-55    & 21.74$-$0.67 & 1995 Jul 15 & 138       & 1800 \\
          &              &             & 57        & 1800 \\
M~4-14    & 43.09$-$3.04 & 1996 Aug 07 & 40        & 1800 \\
          &              &             & 87        & 1800 \\
          &              &             & 153       & 600  \\
WeSb~4    & 31.91$-$0.31 & 1995 Jul 14 & 69        & 1800 \\
          &              &             & 69        & 1800 \\
          &              &             & 159       & 1800 \\

\enddata
\end{deluxetable}

\clearpage

\begin{deluxetable}{lcccccc}
\tabletypesize{\scriptsize}
\tablenum{2}
\tablewidth{0pt}
\tablecaption{Results of Spatio-Kinematical Modeling\label{results}}
\tablehead{
\multicolumn{1}{c}{PN Name}&
\multicolumn{1}{c}{\ha size\tablenotemark{1}}&
\multicolumn{1}{c}{$v_p$} &
\multicolumn{1}{c}{$v_e$} &
\multicolumn{1}{c}{Kin. Age}&
\multicolumn{1}{c}{Incl. Angle}&
\multicolumn{1}{c}{$\gamma$}\\
\multicolumn{1}{c}{}&
\multicolumn{1}{c}{[\arcsec]}&
\multicolumn{1}{c}{[\kms]} &
\multicolumn{1}{c}{[\kms]} &
\multicolumn{1}{c}{[yr]}&
\multicolumn{1}{c}{[\degree]}&
\multicolumn{1}{c}{}}
\startdata
Hen\,2-428 &   63$\times$18   & 80      &  16    &  2400     &  $-$75       & 1 \\
                 &                            & [73,85] & [14,17]& [2400,2500] & $[-74,-76]$& [0.9,1.1]\\
Hen\,2-437&   45$\times$4.6  &    ...        & (5)    &     ...      &  $-$89       & 20 \\
                &                       & [50,100]   & $<10$& [750,2000]    & $[-88,-90]$& [10,45] \\
K\,3-46   &   81$\times$36  & 18         & 3      & 11000       & $-$70        & 0.9 \\
                &                       & [22,32]    & [1,4]  & [7000,12000]& [69,72]   & [0.6,1.2]\\
K\,3-58   &   23.0$\times$12.7  & 38    & 12     & 1800        & $-60 (-57)$  & 2.5 (2.2) \\
               &                        & [32,42]    &[10,15] & [1500,2200] & $[-55,-63)]$& [2.1,2.7]\\
M\,1-75   &   69$\times$23\tablenotemark{2}  & 55         & 8      & 2700        & 87         & 6 \\
                &                       & [45,65]    & [6,10] & [2300,3400] & [86,88]   & [5.5,6.5] \\
M\,1-75\tablenotemark{3}  & ...  & 45         & 12     & 2400        & 65         &  5.5 \\
               &                        & [25,50]    & [6,13] & [2000,5200] & [60,70]   &  [5.3,5.7] \\
M\,2-48  &   42.6$\times$5.7\tablenotemark{4}   &   100      & 10     & 1160        & $-$79        & 8 \\
               &                        & [75,120]   & [8,11] & [800,1600]  & $[-73,-80]$& [7,9] \\
M\,3-55   &   11.1$\times$8.2  & 19.5       & 6      & 1800        & 40         & 0.6 \\
                &                       & [17.5,23.5]& [4,8]  & [1000,2500] & [30,50]   & [0.4,0.9]\\
M\,4-14   &   27$\times$8.5  & 65         &  11    & 1500        &  38        & 5 \\
                &                       & [44,76]    & [7,11] & [1400,2500] &  [37,41]  & [4.7,5.2] \\
WeSb\,4   &   77$\times$18\tablenotemark{4}  & 95         & 14     & 3400        & 50         &  6 \\
                 &                      & [80,110]   & [13,17]& [2750,3700] &
                                       [48,53]   & [5.8,6.2] \\
\enddata
\tablenotetext{1}{The size is of the lobes and waist}
\tablenotetext{2}{Ring size 24.0$\times$13.1}
\tablenotetext{3}{For the smaller pair of lobes}
\tablenotetext{4}{Size in the [N II] image}

\end{deluxetable}
\clearpage

\begin{deluxetable}{lccccccc}
\tabletypesize{\scriptsize}
\tablenum{3}
\tablewidth{0pt}
\tablecaption{Estimated Kinematical Ages\label{ages}}
\tablehead{
\multicolumn{1}{c}{PN Name}&
\multicolumn{1}{c}{$\begin{array}{c}
V_{\rm r}\\
(\kms) \\
\end{array} $} &
\multicolumn{1}{c}{$\begin{array}{c}
{\rm 1~kpc~Kinematic}\\
{\rm Age}\\
{\rm (yr)}\\
\end{array}$} &
\multicolumn{1}{c}{$\begin{array}{c}
{\rm Statistical}\\
{\rm Distance}\\
{\rm (Kpc)}\\
\end{array} $} &
\multicolumn{1}{c}{$\begin{array}{c}
{\rm Reference}\\
\end{array} $} &
\multicolumn{1}{c}{$\begin{array}{c}
{\rm Rotation}\\
{\rm Curve~~Dist.}\\
{\rm (Kpc)}\\
\end{array} $}&
\multicolumn{1}{c}{$\begin{array}{c}
{\rm Adopted}\\
{\rm Distance}\\
{\rm (Kpc)}\\
\end{array} $} &
\multicolumn{1}{c}{$\begin{array}{c}
{\rm Estimated}\\
{\rm Age}\\
{\rm (yr)}\\
\end{array} $ }}
\startdata
Hen\,2-428 & 72 & 2400 & 2.7 & CaKa71     & --& 2.2 & 5280 \\
          &    &      &1.7  & Ma84        &          &     &      \\
K\,3-46    & 66 & 9000 &2.15 & CaKa71     &--&2.2 & 19350 \\
K\,3-58    & 21 & 1800 &  6.6& CaKa71    & --&6.2 & 11070 \\
          &    &      & 5.7  &Ma84    &          &     &       \\
M\,1-75    &  7 & 2700 &2.6-3.7 &CaKa71  & 5.3      & 3.4& 9099 \\
          &    &      &    3.9& Ca76    &          &     &      \\
          &    &      &    2.8& Ac78    &          &     &      \\
          &    &      &    3.21& Da81   &          &     &       \\
          &    &      &    2.40 &AGNR84 &          &     &\\
          &    &      &    7.0 &Ma84    &          &     &\\
          &    &      &    3.89&CKs91  &          &     &\\
M\,2-48    & 16 & 1160 &  3.42 &CaKa71   &        & 4.2 & 4872 \\
          &    &      & 1.6 &Ma84       & 1.5-7.7&     &\\
          &    &      & 6.97 &CKs91     &       &     &\\
M\,3-55    & 30 & 1800 & 3.56 &CaKa71    & 2.8      & 2.8 & 5040 \\
          &    &      &       1.4 &Ma84 &          &     &\\
          &    &      &  3.13 &Ac78     &          &     &\\
M\,4-14    & 49 & 1500 &  3.7 &CaKa71    &  3.7     & 3.7 & 5550 \\
          &    &      &      1.6 &Ma84  &          &     &\\
          &    &      &      2.97 &Ac78 &          &     &\\
          &    &      &     6.69 &CKs91 &          &     &\\
WeSb\,4    & 69 & 3400 & -        & -      & 4.7      &  4.7& 15980 \\

\enddata
\caption[ ]{References: [CaKa71] - Cahn  \& Kaler (1971);
  [Ma84] - Maciel(1984); [Ca76] - Cahn (1976); [Ac78] -
  Acker (1978); [Da82] - Daub (1982); [\emph{AGNR84}] -
	Amnuel et al. (1984); [CKs91] - Cahn et al (1991)} 
\end{deluxetable}

\end{document}